\documentclass[11pt,a4paper]{article}
\pdfoutput=1
\usepackage{jheppub}
\usepackage{epsfig,simplewick,xcolor,amsmath}
\usepackage{fancybox}

\def \Tr{\mbox{Tr\,}}

\newcommand{\be}{\begin{equation}}
\newcommand{\bea}{\begin{eqnarray}}
\newcommand{\ee}{\end{equation}}
\newcommand{\eea}{\end{eqnarray}}
\begin{document}
\title{From Symmetry to Structure: Gauge-Invariant Operators in Multi-Matrix Quantum Mechanics}
\author[a,b]{Robert de Mello Koch\footnote{robert@zjhu.edu.cn},}
\affiliation[a]{School of Science, Huzhou University, Huzhou 313000, China}
\affiliation[b]{Mandelstam Institute for Theoretical Physics, School of Physics, University of the Witwatersrand, Private Bag 3, Wits 2050, South Africa}
\author[c]{Minkyoo Kim\footnote{Corresponding author; mkim@sogang.ac.kr}}
\affiliation[c]{Center for Quantum Spacetime (CQUeST), Sogang University, Seoul 04107, Korea}
\author[d]{Hendrik J.R. Van Zyl\footnote{hjrvanzyl@gmail.com}}
\affiliation[d]{The Laboratory for Quantum Gravity \& Strings,
Department of Mathematics \& Applied Mathematics,
University of Cape Town, Cape Town, South Africa}

\date{}
\abstract{Recently the algebraic structure of gauge-invariant operators in multi-matrix quantum mechanics has been clarified: this space forms a module over a freely generated ring. The ring is generated by a set of primary invariants, while the module structure is determined by a finite set of secondary invariants. In this work, we show that the number of primary invariants can be computed by performing a complete gauge fixing, which identifies the number of independent physical degrees of freedom. We then compare this result to a complementary counting based on the restricted Schur polynomial basis. This comparison allows us to argue that the number of secondary invariants must exhibit exponential growth of the form $e^{cN^2}$ at large $N$, with $c$ a constant.}
\maketitle

\section{Introduction}

The AdS/CFT correspondence \cite{Maldacena:1997re,Witten:1998qj,Gubser:1998bc} is the remarkable equivalence between Type IIB string theory on asymptotically $\text{AdS}_5 \times S^5$ spacetime and $\mathcal{N}=4$ super Yang-Mills theory in four dimensions. The duality provides a non-perturbative definition of quantum gravity in AdS space in terms of a well-understood, conformally invariant gauge theory. The strength of gravitational interactions in the bulk is controlled by $1/N$, so that the large-$N$ limit of the Yang-Mills theory corresponds to a weakly coupled, semiclassical gravity regime.

An early and compelling piece of evidence for the correspondence is the matching between the Kaluza--Klein spectrum of Type IIB supergravity on $S^5$ and the spectrum of chiral primary operators in the boundary gauge theory \cite{Witten:1998qj}. In this dictionary, single-trace, gauge-invariant operators built from elementary fields correspond to the perturbative modes of the bulk string theory. At large $N$ the correlation functions of single trace operators reproduce the free Fock space of supergravity with the number of traces in the CFT mapping into particle number in AdS.

Recent advances, drawing on methods from invariant theory, have yielded a more structured understanding of the space of gauge-invariant operators at finite $N$ \cite{deMelloKoch:2025ngs}. Specifically, in the setting of multi-matrix quantum mechanics it has been shown that this space admits a Hironaka decomposition: it is generated by a set of \emph{primary invariants} $\{O_A\}$, which act freely, together with a set of \emph{secondary invariants} $\{\eta_a\}$, which appear linearly. The general gauge-invariant operator takes the form \cite{deMelloKoch:2025ngs}
\bea
O_1^{n_1} O_2^{n_2} \cdots O_{N_p}^{n_{N_p}}\, \eta_a,
\eea
where $N_p$ denotes the number of primary invariants, and $a$ labels the finite set of secondary invariants. This decomposition reflects the fact that the algebra of invariants at fixed $N$, is a module of finite rank over the free ring generated by the primary invariants. It is striking that the entire space of gauge-invariant operators organizes itself into such a remarkably simple and structured form.

From the perspective of holography, this structure acquires a compelling interpretation. The primary invariants, which generate the ring freely, naturally correspond to the perturbative degrees of freedom in the bulk -- those captured by single-string or supergravity modes. It is notable that the primary invariants are single trace operators so that this interpretation is a natural extension of the duality between supergravity modes and single trace operators. The secondary invariants, by contrast, are more subtle. They appear in precisely the correct way to encode non-trivial gravitational states or backgrounds within the AdS/CFT framework.

Our focus in this paper is on quantum mechanical systems. There are a number of interesting quantum mechanical systems that have gravity duals. By considering $S$-wave and related truncations of ${\cal N}=4$ super Yang-Mills theory~\cite{Kim:2003rza,Okuyama:2002zn,Ishiki:2006rt,Harmark:2014mpa}, one is able to isolate holographic sectors with a clear bulk dual. These truncations lead to quantum mechanical systems. Finite-$N$ effects also play a central role in the description of black hole microstates, particularly in the discovery of the fortuity mechanism~\cite{Chang:2022mjp,Choi:2022caq,Choi:2023znd,Chang:2023zqk,Choi:2023vdm,Chang:2024zqi}: the microstates of $\frac{1}{16}$-BPS black holes fall into two classes, monotone and fortuitous. Monotone states remain BPS at all $N$, while fortuitous states lose their BPS nature above a critical $N$. In the latter case, BPS saturation depends crucially on trace relations. The relevant cohomologies are simpler to study in the BMN matrix model~\cite{Berenstein:2002jq} truncation of the classical field theory.

The number of primary invariants is given by the Krull dimension of the algebra of gauge invariants. For the multi-matrix quantum mechanics of $d$ matrices the Krull dimension is given by $1+(d-1)N^2$. In Section \ref{CountPrim} we explain how the number of primary invariants can be computed by performing a complete gauge fixing of the theory and counting the number of degrees of freedom that remain. We explain how to reproduce the Krull dimension for multi-matrix models and then go onto to predict the number of primary invariants in a theory of bifundamental fields. In a theory with $f$ bifundamental fields we find that the number of primary invariants is given by
\bea
N_p&=&2(f-1)N^2+1\, .
\eea
This formula is confirmed in the appendices where we compute a number of Molien-Weyl partition functions explicitly. The structure of these partition functions reflects both the number of primary and secondary invariants.

The number of secondary invariants is far more difficult to compute. If these do correspond to non-trivial states in the theory, and since we expect the holographic description of multi-matrix quantum mechanics to include black hole solutions, they should account for the black hole microstates. Since black holes have an enormous ($\propto N^2$) entropy this predicts an enormous number ($\sim e^{cN^2}$ with $c=O(1)$) of secondary invariants. In \cite{deMelloKoch:2025ngs} numerical evidence in favour of this dramatic growth was presented. In Section \ref{CountSec} we will manage to give a rigorous argument that establishes this growth.

\section{Counting Primary Invariants}\label{CountPrim}

In this section, we explain how to determine the number of primary invariants associated with a given collection of fields. Our approach begins by performing a complete gauge fixing: we use the gauge symmetry to bring the field configurations into a canonical form, setting as many components as possible to zero. This procedure defines a slice through the space of fields that intersects each gauge orbit exactly once. Mathematically, this corresponds to taking a quotient
\bea
\text{Configuration space} \big/ \text{Symmetry group}&\equiv& {\cal X}\big/G\,,
\eea
which defines the moduli space of gauge-inequivalent configurations.

Invariant theory provides a complementary algebraic formulation of this reduction. Instead of selecting representatives of orbits, it constructs polynomial functions -- \emph{invariants} -- that are constant along each orbit. These invariants generate a finitely generated ring whose structure encodes the geometry of the quotient space. In particular, the number of algebraically independent primary invariants equals the \emph{Krull dimension} of the invariant ring, which matches the dimension of the moduli space ${\cal X}/G$.

In summary, gauge fixing defines \emph{local coordinates} on the orbit space by choosing representatives, while primary invariants furnish \emph{global, algebraic coordinates} that characterize orbits intrinsically. Both approaches yield the same count of physical degrees of freedom, but from different perspectives: gauge fixing is a geometric reduction, while invariant theory offers an algebraic one.

In the sections that follow, we apply this logic to compute the number of primary invariants in multi-matrix models and in theories with bifundamental fields. A similar strategy has already been used successfully in the case of pure vector models~\cite{deMelloKoch:2025cec}.

\subsection{Multiple Matrices}

We consider a theory of $d$ Hermitian $N \times N$ matrices, denoted $X^a$, each transforming in the adjoint representation of the unitary group $U(N)$,
\bea
X^a \mapsto U X^a U^\dagger, \qquad a = 1, 2, \dots, d.
\eea
First, we use this symmetry to simplify any given configuration. Specifically, use the $U(N)$ symmetry to diagonalize the first matrix $X^1$, yielding
\bea
X^1 = \begin{bmatrix}
\lambda_1 & 0 & 0 & \cdots & 0 \\
0 & \lambda_2 & 0 & \cdots & 0 \\
0 & 0 & \lambda_3 & \cdots & 0 \\
\vdots & \vdots & \vdots & \ddots & \vdots \\
0 & 0 & 0 & \cdots & \lambda_N
\end{bmatrix}.\label{canformat}
\eea
This gauge choice does not completely exhaust the $U(N)$ symmetry. The residual symmetry consists of diagonal unitary matrices that preserve the diagonal structure of $X^1$. These residual transformations are given by
\bea
U = \begin{bmatrix}
e^{i\phi_1} & 0 & 0 & \cdots & 0 \\
0 & e^{i\phi_2} & 0 & \cdots & 0 \\
0 & 0 & e^{i\phi_3} & \cdots & 0 \\
\vdots & \vdots & \vdots & \ddots & \vdots \\
0 & 0 & 0 & \cdots & e^{i\phi_N}
\end{bmatrix}.\label{residgt}
\eea
The residual phases $\phi_i$ can be fixed to simplify the remaining matrices. Consider how such a transformation acts on $X^2$, with $m_{ab}$ denoting its components and $\phi_{ab} \equiv \phi_a - \phi_b$:
\bea
U X^2 U^\dagger = 
\begin{bmatrix}
m_{11} & e^{-i\phi_{21}} m_{12} & e^{-i\phi_{31}} m_{13} & \cdots & e^{-i\phi_{N1}} m_{1N} \\
e^{i\phi_{21}} m_{21} & m_{22} & e^{-i\phi_{32}} m_{23} & \cdots & e^{-i\phi_{N2}} m_{2N} \\
e^{i\phi_{31}} m_{31} & e^{i\phi_{32}} m_{32} & m_{33} & \cdots & e^{-i\phi_{N3}} m_{3N} \\
\vdots & \vdots & \vdots & \ddots & \vdots \\
e^{i\phi_{N1}} m_{N1} & e^{i\phi_{N2}} m_{N2} & e^{i\phi_{N3}} m_{N3} & \cdots & m_{NN}
\end{bmatrix}.
\eea
Notice that all the induced phases depend only on differences $\phi_a - \phi_b$. Consequently, a common shift $\phi_a \to \phi_a + \alpha$ leaves all matrices invariant. Hence, of the $N$ residual phases $\phi_a$, only $N - 1$ are physically meaningful and can be used to further reduce the degrees of freedom.

Since $X^2$ is Hermitian, its matrix elements satisfy $m_{ab} = m_{ba}^*$. The off-diagonal entries thus contain non-trivial phases. We can use the $N - 1$ phase differences $\phi_a - \phi_{a+1}$, for $a = 1, 2, \dots, N-1$, to eliminate the phases of $m_{a\,a+1}$, setting each to be real. These are the additional fields we can fix using the residual symmetry.

We now count the remaining degrees of freedom. There are $(d - 1)N^2$ matrix elements in the fields $X^a$ for $a = 2, \dots, d$, and $N$ eigenvalues in the diagonal matrix $X^1$. After fixing $N - 1$ phases in \( X^2 \) as described above, we find
\bea
N_P&=&(d - 1)N^2 + N - (N - 1) = 1 + (d - 1)N^2\label{NP}
\eea
independent parameters. This correctly counts the number of \textit{primary invariants} derived from a system of $d>1$ $N\times N$ Hermitian matrices, and agrees with computation of the Krull dimension performed in \cite{herstein}. This result is intuitively appealing: since the $X^a$ are in the adjoint representation, they are invariant under the $U(1)$ in $U(N)\simeq SU(N)\times U(1)$, so that the effective gauge group is $SU(N)$ of dimension $N^2-1$. Thus, gauge fixing eliminates $N^2-1$ degrees of freedom from the total $dN^2$ to leave (\ref{NP}).

For the case that $d=1$, the canonical form of the field configuration defining the gauge fixing is (\ref{canformat}). In this case the residual gauge redundancy spelled out in (\ref{residgt}) does not set any further components to zero and the number of primary invariants for a single matrix is
\bea
N_P&=&N
\eea
matching the number of eigenvalues.

\subsection{Bifundamental fields}

We now turn to the task of counting the number of primary invariants in a system of $f$ bifundamental fields. In this example, the gauge group is $U(N)\times U(N)$. For notational simplicity, we denote the bifundamentals transforming in the $(\mathbf{N}, \overline{\mathbf{N}})$ representation of $U(N) \times U(N)$ by $\Gamma^A \equiv \Gamma_{12}^A$, and those in the $(\overline{\mathbf{N}}, \mathbf{N})$ by $\bar{\Gamma}^A \equiv \Gamma_{21}^A$. Under the gauge group, these transform as
\bea
\Gamma^A \to U \Gamma^A V^\dagger, \qquad \bar{\Gamma}^A \to V \bar{\Gamma}^A U^\dagger\qquad A=1,2,\cdots,f.
\eea
We begin by counting the number of independent field components prior to gauge fixing. A positive-definite real action for the bifundamentals takes the form ($A$ is summed from 1 to $f$)
\bea
S = \int dt \left( \Tr\left(\partial_t \Gamma^A \partial_t \bar{\Gamma}^A\right) - \Tr\left(\Gamma^A \bar{\Gamma}^A\right) \right),
\eea
with the identification $\bar{\Gamma}^A = (\Gamma^A)^\dagger$. Consequently, each $\Gamma^A$ is a general complex $N \times N$ matrix, while $\bar{\Gamma}^A$ is determined from $\Gamma^A$ by Hermitian conjugation. A complex matrix has $2N^2$ real degrees of freedom, so for each flavor index $A$, there are $2N^2$ field components. For a system with $f$ flavors, the total number of real field components is $2fN^2$.

We now utilize the gauge symmetry to fix redundant degrees of freedom. The singular value decomposition allows us to write the first bifundamental as
\bea
\Gamma^1 = U^\dagger D V,
\eea
where $U, V \in U(N)$ are unitary matrices and $D$ is a diagonal matrix with non-negative real entries. Using the gauge symmetry, we can bring $\Gamma^1$ into the canonical diagonal form
\bea
\Gamma^1 = D.\label{gaugefixed}
\eea
This fixes $2N^2 - N$ of the original $2N^2$ components in $\Gamma^1$, leaving only the $N$ non-negative entries on the diagonal.

However, this gauge fixing is not unique. The residual symmetry consists of transformations $U, V$ that are diagonal and equal unitary matrices,
\bea
U = \mathrm{diag}(e^{i\phi_1}, \dots, e^{i\phi_N}) = V,\label{residuals}
\eea
that preserve the diagonal and real form of $D$. These $N$ phases are subject to an overall $U(1)$ redundancy (a common shift), so we can use $N - 1$ residual parameters to further simplify the remaining fields.

In particular, we apply these residual transformations to $\Gamma^2$, which remains unfixed. Since $\Gamma^2$ is a complex matrix, the phases in its off-diagonal entries are complex. As in the case of a single Hermitian matrix, we can use $N-1$ relative phase rotations to eliminate the imaginary parts of $N-1$ off-diagonal entries. For example, we may choose to set the phases of the entries $(\Gamma^2)_{a, a+1}$ to zero for $a=1,\dots,N-1$.

Thus, the number of independent real field components remaining after fully exhausting the gauge symmetry is
\bea
N_P&=&2(f - 1)N^2 + 1.
\eea
This provides our prediction for the number of primary invariants in the model of $f$ bifundamental fields, when $f>1$.

When $f=1$ the gauge fixing is given by (\ref{gaugefixed}), and the residual gauge transformations (\ref{residuals}) are not able to set any further field components to zero. Thus, for $f=1$ the number of primary invariants is given by
\bea
N_P&=&N
\eea

\section{Counting of Secondary Invariants}\label{CountSec}

In this section, we consider the growth of the number of secondary invariants in the two-matrix model. This setting is particularly tractable because the complete spectrum of energy eigenstates can be explicitly constructed using restricted Schur polynomials, which provide a natural orthogonal basis adapted to the underlying gauge symmetry. By comparing two distinct but complementary formulations of the partition function -- the Hironaka decomposition, which organizes the invariant ring as a free module over the ring of primary invariants, and the combinatorial form arising from the restricted Schur polynomial basis -- we obtain a nontrivial bound on the number of secondary invariants. This analysis leverages known asymptotic estimates for the growth of Littlewood–Richardson coefficients, which govern the multiplicities in tensor product decompositions of representations. The outcome is striking: the number of secondary invariants grows as $e^{cN^2}$ for some constant $c$. This result highlights the combinatorial complexity that emerges from non-planar degrees of freedom even in a seemingly simple two-matrix model.

Recall that the partition function for the singlet sector of decoupled matrix oscillators, one oscillator for each of $d$ matrices, is given by
\bea
Z(x)&=&\frac{1+\sum_i c^s_i x^i}{\prod_{j}(1-x^j)^{c^p_j}}\label{PartFunc}
\eea
We know that the number of primary invariants is given by
\bea
1+(d-1)N^2&=&\sum_{j}c^p_j
\eea
while the number of secondary invariants, denoted $N_s$, is given by
\bea
N_s&=&1+\sum_i c^s_i
\eea
In this section we study $N_s$. Although we are not able to compute $N_s$ with any precision, we can show that it must grow as the exponential of $N^2$. Although we focus on the case of $d=2$ it is clear that this establishes a growth as the exponential of $N^2$ for $d>2$ as well.


We want to construct the complete set of eigenstates of the two matrix oscillator
\bea
H&=&{1\over 2}\Tr(\Pi^1\Pi^1)+{1\over 2}\Tr(X^1X^1)+{1\over 2}\Tr(\Pi^2\Pi^2)+{1\over 2}\Tr(X^2X^2)
\eea
\bea
[\Pi^a_{ij}(t),X^b_{kl}(t)]&=&-i\delta^{ab}\delta_{jk}\delta_{il}
\eea
The eigen-problem of this oscillator is solved, as usual, in terms of creation and annihilation operators, where
\bea
A^{a}_{ij}=X^a_{ij}-i\Pi^a_{ij}
\eea
The eigenstates are given by polynomials in the creation operators of a fixed degree. We can refine this further by grading with respect to degree in $A^{1\,\dagger}$ and degree in $A^{2\,\dagger}$ separately. Any multi-trace operator, constructed from $n$ $A^{1\,\dagger}$ fields and $m$ $A^{2\,\dagger}$ fields can be written as
\bea
{\rm Tr}\left(\sigma {A^{1\,\dagger}}^{\otimes n}\otimes {A^{2\,\dagger}}^{\otimes m}\right)&=&{A^{1\,\dagger}}^{i_1}_{i_{\sigma(1)}}\cdots {A^{1\,\dagger}}^{i_n}_{i_{\sigma(n)}}{A^{2\,\dagger}}^{i_{n+1}}_{i_{\sigma(n+1)}}\cdots {A^{2\,\dagger}}^{i_{n+m}}_{i_{\sigma(n+m)}}\label{eigenop}
\eea
where $\sigma\in S_{n+m}$. The complete set of such operators creates the complete set of states of energy $n+m$ when they are applied to the ground state $|0\rangle$. A basis for these operators is given by the restricted Schur polynomials \cite{Bhattacharyya:2008rb} (see \cite{deMelloKoch:2024sdf} for a pedagogical introduction)
\bea
\chi_{R,(r,s)\alpha\beta}({A^{1\,\dagger}},{A^{2\,\dagger}})={1\over n!m!}\sum_{\sigma\in S_{n+m}}\chi_{R,(r,s)\alpha\beta}(\sigma){\rm Tr}\left(\sigma {A^{1\,\dagger}}^{\otimes n}\otimes {A^{2\,\dagger}}^{\otimes m}\right)
\eea
The labels for the restricted Schur polynomial is a Young diagram $R$ with $n+m$ boxes specifying a representation $S_{n+m}$, a Young diagram $r$ with $n$ boxes specifying a representation of $S_n$, a Young diagram $s$ with $m$ boxes specifying a representation of $S_m$ and a pair of multiplicity labels. The multiplicity labels range from $1$ to $f_{Rrs}$ where $f_{Rrs}$ is the Littlewood-Richardson number counting how many times the U($N$) representation $R$ appears in the product of U($N$) representations $r$ and $s$. To see that the restricted Schur polynomials provide a complete basis, use the completeness identity for restricted characters \cite{Bhattacharyya:2008xy}
\bea
\sum_{\rho\in S_{n+m}}\chi_{R,(r,s)\alpha\beta}(\rho^{-1})\chi_{T,(t,u)\gamma\delta}(\rho)&=&(n+m)!\delta_{RT}\delta_{rt}\delta_{su}\delta_{\gamma\beta}\delta_{\alpha\delta}
\eea
\bea
\sum_{R,(r,s)\alpha\beta}\chi_{R,(r,s)\alpha\beta}(\tau)\chi_{R,(r,s)\alpha\beta}(\sigma)&=&(n+n)!{{\rm dim}_r {\rm dim}_s\over {\rm dim}_R}\delta_{[\sigma]_r[\tau]_r}
\eea
where $\delta_{[\sigma]_r[\tau]_r}$ is 1 if $[\sigma]_r$ and $[\tau]_r$ belong to the same restricted conjugacy class and it is zero otherwise. If two permutations $\rho_1$ and $\rho_2$ are related as
\bea
\rho_1 &=&\sigma^{-1}\rho_2\sigma\qquad\qquad\rho_1,\rho_2\in S_{n+m}\quad \sigma\in S_n\times S_m
\eea
then they belong to the same restricted conjugacy class. We can now write any operator of the form (\ref{eigenop}) in terms of restricted Schur polynomials, using the completeness of restricted characters, as follows
\bea
{\rm Tr} (\sigma {A^{1\,\dagger}}^{\otimes n}{A^{2\,\dagger}}^{\otimes m})&=&\sum_{T,(t,u)\alpha\beta}{d_T n! m!\over d_t d_u (n+m)!}\chi_{(t,u)\alpha\beta}(\sigma^{-1})\chi_{T,(t,u)\beta\alpha}({A^{1\,\dagger}},{A^{2\,\dagger}})\cr
&&
\eea
Thus, the partition function (\ref{PartFunc}) can equivalently be computed from the counting of restricted Schur polynomials, which leads to the formula
\bea
Z(x)&=&\sum_{\substack{R\vdash n+m\\ l(r)\le N}}\sum_{r\vdash n}\sum_{s\vdash m}\,\, (f_{Rrs})^2\,\, x^{n+m}\cr\cr
&=&\cdots +c^T_{\alpha N^2} x^{\alpha N^2}+\cdots
\eea
We have indicated where our interest is: in the coefficient $c^T_{\alpha N^2}$ of the term $x^{\alpha N^2}$ which counts the number of operators of degree $\alpha N^2$. Since this counts the total contribution at degree $\alpha N^2$, the superscript ``$T$'' is for total. We have in mind that $\alpha$ is held fixed as we take $N$ large. This formula is useful because the asymptotic behaviour of the Littlewood-Richardson number needed to estimate the behaviour of the coefficient of $x^q$ with $q$ of order $N^2$ is known \cite{Stanley1,Stanley2,PakPanova}. If $R$ has $n$ boxes, $r$ has $k$ boxes and $s$ has $n-k$ boxes, when $n$, $k$ and $n-k$ are all large, the largest Littlewood-Richardson coefficient is given by 
\bea
f_{Rrs}=e^{{n\over 2}\log 2+O(\sqrt{n})}\label{maxLR}
\eea
We can produce an underestimate of the coefficient $c^T_{\alpha N^2}$ by summing only the term corresponding to the largest Littlewood-Richardson coefficient. The result is that
\bea
c^T_{\alpha N^2}>e^{\alpha N^2 \log 2+O(N)}\label{underest}
\eea
The result (\ref{maxLR}) may be modified by finite $N$ constraints. This implies that the Young diagram $R$ must not have more than $N$ rows, It is therefore necessary to inquire about the shape of the Young diagrams $R$, $r$ and $s$ that attain the maximum Littleoowd-Richardson numbers. It turns out, that all three Young diagrams must take the Vershik-Kerov-Logan-Shepp (VKLS) shape\cite{VK1,VK2,LS}. These Young diagrams have a triangular shape and are not cut off by finite $N$ effects for Young diagrams with $\alpha N^2$ boxes, as long as $\alpha$ is not too large. Young diagrams with even more boxes take on the truncated VKLS shape \cite{LS}.

A comment is in order. Solving for the eigenfunctions of the multi-matrix harmonic oscillator, with oscillators for different matrices decoupled, is easily solved using creation and annihilation operators. Eigenstates are then given by any multi-trace polynomials in the creation oscillators, acting on the vacuum and the energy is just the degree of the polynomial. This leads to states with huge degeneracies and it becomes difficult to manage these large degeneracies. By working with restricted Schur polynomials, we organize this enormous set of states into an orthogonal basis. Indeed, with a little work we find the eigenfunctions
\bea
|R,(r,s)\alpha\beta\rangle&\equiv&\chi_{R,(r,s)\alpha\beta}(A^{1\,\dagger},A^{2\,\dagger})|0\rangle
\eea
which have the following inner products
\bea
\langle T,(t,u)\gamma\delta|R,(r,s)\alpha\beta\rangle&=&\delta_{RS}\delta_{rt}\delta_{su}\delta_{\gamma\alpha}\delta_{\beta\delta}{f_R{\rm hooks}_R\over {\rm hooks}_r{\rm hooks}_s}
\eea
with $f_R$ the product of the factors for each box in Young diagram $R$ and ${\rm hooks}_l$ the product of the hook lengths of each box in Young diagram $l$. Exactly the same analysis could be carried with the bases introduced in \cite{BHR}.

Next rewrite (\ref{PartFunc}) as
\bea
Z(x)&=&\frac{1}{\prod_{j}(1-x^j)^{c^p_j}}(1+\sum_i c^s_i x^i)\,\,\equiv\,\, Z_p(x)(1+\sum_i c^s_i x^i)
\eea
We are going to estimate the coefficient $d_{\alpha N^2}$ defined by
\bea
Z_p(x)&=&\frac{1}{\prod_{j}(1-x^j)^{c^p_j}}\,\,=\,\,\sum_q d_q x^q\cr\cr
&=&\cdots+ d_{\alpha N^2}\, x^{\alpha N^2}+\cdots
\eea
We know that there is a factor in the denominator for each primary invariant, so that there are $N^2+1$ factors in total. Separating $Z_p(x)$ into a product of $N^2+1$ factors, each factor can be expanded in an infinite series. Each such factor is an infinite sum of monomials, each with coefficient 1
\bea
 {1\over 1-x^i}&=&1+x^i +(x^i)^2+(x^i)^3+(x^i)^4+\cdots
\eea
The coefficient $d_{\alpha N^2}$ is simply counting how many ways we can partition $\alpha N^2$ into parts, with the parts used in the partition process defined by the $c^p_j$: there are $c_p^j$ different parts of length $j$. It is difficult to estimate $d_{\alpha N^2}$ since we don't have a complete understanding of the integers $c^p_j$. We  know that $c^p_1=2$ and that there are non-zero values of $c^p_j$ for $j$ increasing from 1 to $O(N)$. These non-zero integers record how many single trace operators of length $j$ are selected as primary invariants. A much simpler problem is to estimate
\bea
\hat{Z}_p(x)&=&\frac{1}{(1-x)^{N^2+1}}\,\,=\,\,\sum_q \hat{d}_q x^q\cr\cr
&=&\cdots+ \hat{d}_{\alpha N^2}\, x^{\alpha N^2}+\cdots
\eea
In this case, $\hat{d}_{\alpha N^2}$ is counting how many ways we can partition $\alpha N^2$ into parts, using $N^2+1$ distinct parts, all of length 1 and we can use each as many times as we wish. It is obvious that
\bea
d_{\alpha N^2}<\hat{d}_{\alpha N^2}
\eea
so we are overestimating $d_{\alpha N^2}$. The point is now that $\hat{d}_{\alpha N^2}$ is easily estimated. First, note that the coefficient $\beta_p$ defined by
\bea
{1\over (1-y)^q}=\cdots+\beta_p y^p+\cdots
\eea
is given by
\bea
\beta_p&=&\frac{(p+q-1)!}{(q-1)!p!}\label{FormFrCff}
\eea
Our goal is to show that, even with an over estimate, $d_{\alpha N^2}$ does not grow fast enough with $N$ to be consistent with (\ref{underest}). Applying this formula we easily obtain
\bea
d_{\alpha N^2}<\hat{d}_{\alpha N^2}&=& e^{((\alpha+1)\log(\alpha+1)-\alpha\log(\alpha))N^2}
\eea
which does not prove an inconsistency with (\ref{underest}), at least for $\alpha \leq 1$. However, this bound is very weak and we can improve it significantly, using some knowledge we have of the $c^p_j$. Specifically, as $N$ increases we know that more and more of the traces of small length are included as primary invariants. For $N\ge 4$ for example, we know that 
\bea
c^p_1&=&2\qquad c^p_3\,\,=\,\,3\qquad c^p_4\,\,=\,\,6
\eea
Motivated by this observation, we can introduce a family of partition functions indexed by an integer $k$ as follows
\bea
\hat{Z}_P(x) & = & \frac{1}{(1 - x)^{C_k}}\frac{1}{(1 - x^{k})^{N^2 + 1 - C_k}}\label{hatPF}
\eea
where
\bea
C_k & = & \sum_{i=1}^{k-1} c^p_i
\eea
The expansion coefficients $\hat{d}_{\alpha N^2}$ continue to bound $d_{\alpha N^2}$ for any $k$, with the bound improving as $k$ is increased. The expansion of the partition function is
\begin{eqnarray}
\hat{Z}_P(x) & = & \sum_{n_1, n_k}  \frac{(n_1+C_k -1)!}{n_1! (C_k - 1)!} \frac{(n_k+ N^2 - C_k)!}{n_k! (N^2 - C_k)!} x^{n_1 + k n_k}    \nonumber \\
& = & \sum_{n, n_k}\sum_{j=0}^{k-1}  \frac{(k n + j +C_k -1)!}{(k n + j)! (C_k - 1)!} \frac{(n_k+ N^2 - C_k)!}{n_k! (N^2 - C_k)!} x^{k (n_k + n) + j}   \nonumber \\
& = & \sum_{M} \sum_{n=0}^M \sum_{j=0}^{k-1}  \frac{(k n + j +C_k -1)!}{(k n + j)! (C_k - 1)!} \frac{(M-n+ N^2 - C_k)!}{(M-n)! (N^2 - C_k)!} x^{k M + j} \label{Zkfunc}
\end{eqnarray}
The sums above may be reorganised in the following convenient way
$$\hat{Z}_P(x) =  \sum_M  \frac{(M+N^2 - C_k)!}{M! (N^2 - C_k)!}  \sum_{j=0}^{k-1} \sum_{n=0}^M  \frac{(k n + j +C_k -1)!}{(k n + j)! (C_k - 1)!} \frac{(-M)_{(n)}}{(C_k-M-N^2)_{(n)}} x^{k M + j}     $$
where $(a)_{(n)}$ denotes the Pochhammer symbol.  The sum over $n$ may be performed explicitly to give a generalized hypergeometric function.  The explicit expression is not too important since we are only interested in its large $N$ behaviour.  
%
%
For $N$ large and $k$ of order $1$, $C_k$ is an $O(1)$ number. We then find that the coefficient of $x^{\alpha N^2}$ scales as
\bea 
e^{\frac{1}{k}((\alpha + k) \log(1 + \frac{\alpha}{k}) - \alpha \log(\frac{\alpha}{k})  )N^2+O(N)}
\eea
Note that the exponential growth with $N^2$ is being set by the final factor on the RHS of (\ref{hatPF}), exactly as it should be. Now, setting $k=15$ this allows us to bound 
\bea
d_{\alpha N^2}<e^{{1\over 15}((15+\alpha)\log (1+{\alpha\over 15})-\alpha\log{\alpha\over 15})N^2+O(N)}\label{weakbound}
\eea
{From the partition function
\bea
Z(x)&=&\sum_q d_q x^q\,\Big(1+\sum_i c_i^s x^i\Big)
\eea
we easily obtain the following formula for $c^T_{\alpha N^2}$
\bea
c_{\alpha N^2}^T&=&d_{\alpha N^2}+\sum_{i=1}^{\alpha N^2}c_i^s d_{\alpha N^2-i}
\eea
Now, we use the fact that the coefficients $d_l$ are strictly increasing with $l$ which follows immediately because the number of ways to partition a positive integer grows as the integer grows\footnote{We have two primary invariants of length 1. Thus, the partitions of $l+1$ include (at least) the partitions of $l$ multiplied by the length 1 primaries. Thus $d_{l+1}>d_l$.}. Consequently we find
\bea
1+\sum_{i=1}^{\alpha N^2}c_i^s &\ge&
1+\sum_{i=1}^{\alpha N^2}c_i^s {d_{\alpha N^2-i}\over d_{\alpha N^2}}\cr\cr
&=&{1\over d_{\alpha N^2}}\left(d_{\alpha N^2}+\sum_{i=1}^{\alpha N^2}c_i^s d_{\alpha N^2-i}\right)\cr\cr
&=&{c^T_{\alpha N^2}\over d_{\alpha N^2}}>{c^T_{\alpha N^2}\over \hat{d}_{\alpha N^2}}\cr\cr
&>&e^{\alpha \log 2\, N^2 - {1\over k}\left((k+\alpha)\log (1+{\alpha\over k}) - \alpha \log{\alpha\over k}\right)N^2}
\eea
Consequently, we obtain the cumulative lower bound
\bea
1+\sum_{i=1}^{\alpha N^2}c_i^s &>&
e^{\Big(\alpha \log 2 - {1\over k}\left((k+\alpha)\log (1+{\alpha\over k}) - \alpha \log{\alpha\over k}\right)\Big)N^2}
\eea
To establish our central claim we need to study the sign of the coefficient of $N^2$ in the exponent and verify that it is positive.}
\vfill\eject

\begin{figure}[h]
\begin{center}
\includegraphics[width=0.75\linewidth]{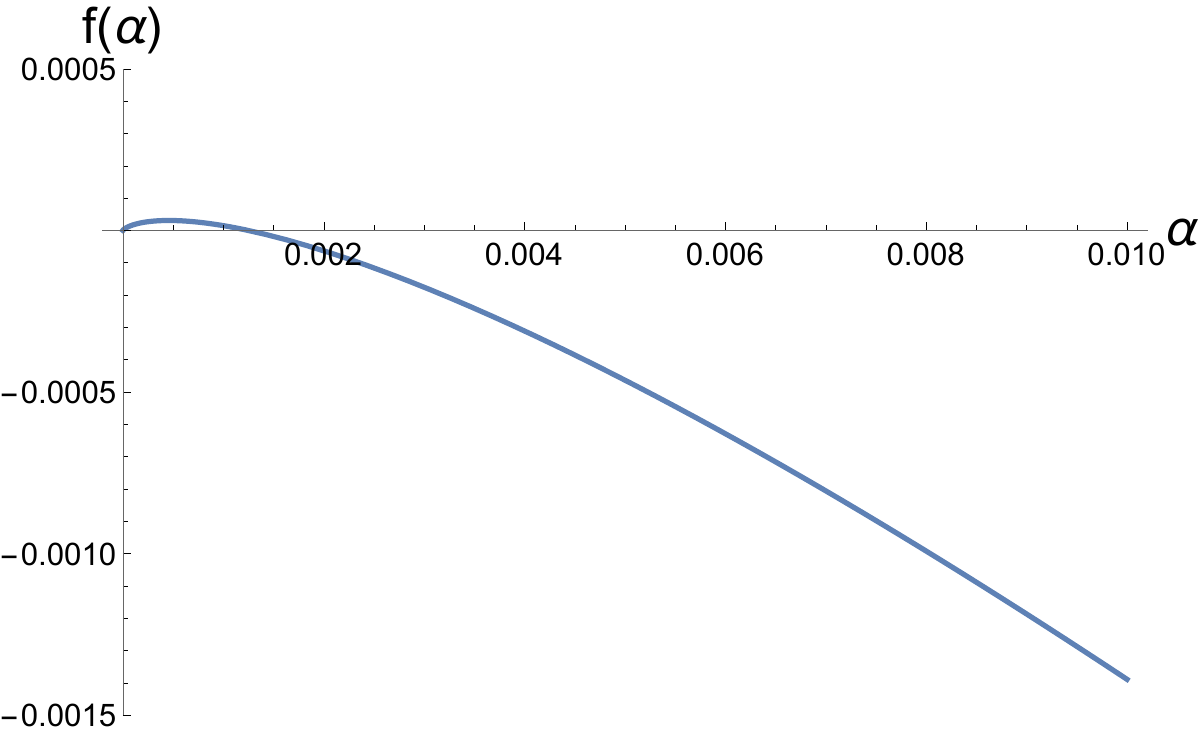}
\end{center}
\caption{{\bf A plot of the function $f(\alpha)={1\over 15}\left((15+\alpha)\log (1+{\alpha\over 15})-\alpha\log{\alpha\over 15}\right)-\alpha\log (2)$.}}
\label{alphafunc}
\end{figure}

From Figure \ref{alphafunc} it is clear that for any choice of $\alpha> \frac{1}{500}$ we have
\bea
\alpha \log 2 > {1\over 15}\left((15+\alpha)\log (1+{\alpha\over 15}) - \alpha \log{\alpha\over 15}\right)
\eea
so that the number of secondary invariants must grow as
\bea
1+\sum_i c^s_i &>& e^{c N^2}
\eea
with $c$ a positive $O(1)$ number, which is the result we wanted to prove.

It is worth making a few comments on the bound (\ref{weakbound}). This bound is very weak. Indeed, we have only used a few values of $c^p_j$ for small values of $j$. By using more and more values of $c^p_j$ we can write more detailed versions of $\hat{Z}_p(x)$ which provide stricter and stricter bounds on $d_{\alpha N^2}$. It is not even clear to us that the bound (\ref{weakbound}) captures the correct dependence on $N$.

\section{Discussion}

There are two important lessons from our results:
\begin{itemize}
\item[1.] The number of primary invariants can be computed by performing a complete gauge fixing of the field configurations.
\item[2.] The number of secondary invariants grows exponentially as $e^{cN^2}$, with $c$ a constant.
\end{itemize}
Both results admit a compelling interpretation.

The connection between gauge fixing and the enumeration of primary invariants underpins a deep structural correspondence. In the matrix model, according to our interpretation outlined in the introduction, working in a given background corresponds to working in the Fock space associated to some secondary operator. In the description employing invariants, the primary invariants provides the coordinates for this space. On the other hand, in quantum mechanics, a set of coordinates on a particular gauge slice is given by the degrees of freedom that remain after fully fixing the gauge. It is satisfying to see the match between the number of primary operators (the Krull dimension of the ring generated by the primaries) and the number of degrees of freedom  counted after a complete gauge fixing.

As we discussed in the introduction, if the secondary invariants correspond to non-trivial states in the theory, because the holographic description of multi-matrix quantum mechanics includes black hole solutions, they must account for the black hole microstates. The enormous entropy of black holes ($\propto N^2$) then predicts an enormous number ($\sim e^{cN^2}$ with $c=O(1)$) of secondary invariants. We now have a proof that the secondary invariants indeed exhibit this growth. Thus, we obtain non-trivial support for the identification of secondary invariants with black hole microstates from purely field-theoretic considerations.

Our analysis has focused on counting. It is interesting to go beyond enumerative results and study how these ideas constrain or reveal aspects of black hole dynamics. For instance, can secondary invariants be used to probe thermalization, scrambling, or chaotic behavior? A dynamical characterization of the role played by secondary invariants would greatly strengthen their proposed interpretation and open a new window onto the microscopic physics of quantum gravity.

\begin{center} 
{\bf Acknowledgements}
\end{center}
The work of RdMK is supported by a start up research fund of Huzhou University, a Zhejiang Province talent award and by a Changjiang Scholar award. The work of MK is supported by NRF grants    RS-2025-25414114, RS-2023-NR077094, RS-2020-NR049598(CQUeST). HJRvZ would like to acknowledge support from the “Quantum Technologies for Sustainable
Development” grant from the National Institute for Theoretical
and Computational Sciences of South Africa (NITHECS).

{\vskip 2.0cm}

\begin{appendix}

\section{Molien-Weyl Functions}

We study a model of $b_1$ bosonic matrices ($\Phi_1^I$) and $b_2$ bosonic matrices ($\Phi_2^I$). There are also $f$ pairs of bifundamental fields $\Gamma_{12}^A$ and $\Gamma_{21}^A$. To keep track of the various fields of the theory, introduce four sets of chemical potentials $\mu_{\Phi_1^I}$, $\mu_{\Phi_2^I}$, $\mu_{\Gamma_{12}^A}$ and $\mu_{\Gamma_{21}^A}$.  Assume all states have unit energy spacing. The partition function is written in terms of the variables
\bea
w_{12}^A=e^{-\beta E_{\Gamma_{12}^A} -\mu_{\Gamma_{12}^A}}\qquad w_{21}^A=e^{-\beta E_{\Gamma_{21}^A} -\mu_{\Gamma_{21}^A}}\cr\cr
z_1^I=e^{-\beta E_{\Phi_1^I}-\mu_{\Phi_1^I}}\qquad z_2^I=e^{-\beta E_{\Phi_2^I}-\mu_{\Phi_2^I}}
\eea
Since we have a free theory the spectrum is additive and the exact partition function is
\bea
Z(w^A,z^I)&=&\prod_{A_1,A_2=1}^f\,\prod_{I_1=1}^{b_1}\,\prod_{I_2=1}^{b_2}\sum_{n_{A_1}=0}^\infty\sum_{n_{A_2}=0}^\infty\sum_{p_{I_1}=0}^\infty\sum_{p_{I_2}=0}^\infty (w^{A_1}_{12})^{n_{A_1}}(w^{A_2}_{21})^{n_{A_2}}(z_1^{I_1})^{p_{I_1}}(z_2^{I_2})^{p_{I_2}}\cr\cr
&&\qquad\times \#(\{n_{A_1},n_{A_2},p_{I_1},p_{I_2}\})
\eea
where $\#(\{n_{A_1},n_{A_2},p_{I_1},p_{I_2}\})$ is the number of singlets that can be constructed using $n_{A_1}$ $\Gamma^{A_1}_{12}$ bifundamental fields, $n_{A_2}$ $\Gamma^{A_2}_{12}$ bifundamental fields, $p_{I_1}$ adjoints of the first $U(N)$ factor and $p_{i_2}$ adjoints of the second $U(N)$ factor. Using characters we can write the number of singlets as an integral over $U(N)\times U(N)$ to obtain
\bea
Z(x^i,y^i,z^I)&=&\int_{U_1(N)} [DU_1]\,\int_{U_2(N)} [DU_2]\,\prod_{A_1,A_2=1}^f\sum_{n_{A_1}=0}^\infty (w^{A_1}_{12})^{n_{A_1}}\chi_{{\rm sym}^{n_{A_1}}(F)}(U_1)
\chi_{{\rm sym}^{n_{A_1}}(\bar{F})}(U_2)\cr\cr
&&\times\,\sum_{n_{A_1}=0}^\infty(w^{A_2}_{21})^{n_{A_2}}\chi_{{\rm sym}^{n_{A_2}}(F)}(U_2)\chi_{{\rm sym}^{n_{A_2}}(\bar{F})}(U_1)\prod_{I_1=1}^{b_1} \sum_{p_{I_1}=0}^\infty (z_1^{I_1})^{p_{I_1}}\chi_{{\rm sym}^{p_{I_1}}(adj)}(U_1)\cr\cr
&&\qquad\times\prod_{I_2=1}^{b_2} \sum_{p_{I_2}=0}^\infty (z_2^{I_2})^{p_{I_2}}\chi_{{\rm sym}^{p_{I_2}}(adj)}(U_2)
\eea
$adj$ stands for the adjoint representation. As always, symmetric products of representations appear above because the fields we are using are bosonic. A useful identity is
\bea
\sum_{n=0}^\infty x^n \chi_{{\rm sym}^{n}(R)}(U)={\rm exp}\left(\sum_{m=1}^\infty {x^m\over m}\chi_R(U^m)\right)
\eea
The partition function becomes
\bea
Z(x^i,y^i,z^I)&=&\int_{U_1(N)} [DU_1]\,\,\int_{U_2(N)} [DU_2]\cr\cr
&&\,\,{\rm exp}\left(\sum_{I_1=1}^{b_1} \sum_{m=1}^\infty {(z^{I_1}_1)^m\over m}\chi_{adj}(U_1^m)+\sum_{I_2=1}^{b_2} \sum_{m=1}^\infty {(z^{I_2}_2)^m\over m}\chi_{adj}(U_2^m)\right)\cr\cr
&&\times{\rm exp}\left(\sum_{A_1=1}^f\sum_{m=1}^\infty{(w_{12}^{A_1})^m\over m}
\chi_F(U_1^m)\chi_{\bar{F}}(U_2^m)+\sum_{A_2=1}^f\sum_{m=1}^\infty{(w_{21}^{A_2})^m\over m}
\chi_{F}(U_2^m)\chi_{\bar{F}}(U_1^m)\right) \nonumber
\eea
The integrand above depends only on the eigenvalues $\varepsilon_{1i},\varepsilon_{2i}$ of $U_1,U_2$. Recall that the eigenvalues of the unitary group obey $|\varepsilon_i|=1$ and $\varepsilon_i^*=\varepsilon_i^{-1}$. Using the explicit expressions for the characters in terms of the eigenvalues
\bea
\chi_{\rm adj}(U_1^{m})&=&\sum_{i,j=1}^N (\varepsilon_{1i})^{m} (\varepsilon_{1j})^{-m}\qquad \chi_{\rm adj}(U_2^{m})\,\,=\,\,\sum_{i,j=1}^N (\varepsilon_{2i})^{m}(\varepsilon_{2j})^{-m}\cr\cr
\chi_{F}(U_1^{m})&=&\sum_{i=1}^N (\varepsilon_{1i})^{m} \qquad\qquad \chi_{F}(U_2^{m})\,\,=\,\,\sum_{i=1}^N (\varepsilon_{2i})^{m}\cr\cr
\chi_{\bar F}(U_1^{m})&=&\sum_{j=1}^N  (\varepsilon_{1j})^{-m}\qquad\qquad \chi_{\bar F}(U_2^{m})\,\,=\,\,\sum_{j=1}^N (\varepsilon_{2j})^{-m}
\nonumber
\eea
and the fact that for an integrand that depends only on the eigenvalues we can replace 
\bea
\int_{U_1(N)} [DU_1] &\to& {1\over N! (2\pi i)^N}\oint \prod_{j=1}^N {d\varepsilon_{1j}\over\varepsilon_{1j}}\Delta(\varepsilon_1)\bar{\Delta}(\varepsilon_1)\cr\cr
\int_{U_2(N)} [DU_2] &\to& {1\over N! (2\pi i)^N}\oint \prod_{j=1}^N {d\varepsilon_{2j}\over\varepsilon_{2j}}\Delta(\varepsilon_2)\bar{\Delta}(\varepsilon_2)
\eea
where the Van der Monde determinants are given by
\bea
\Delta(\varepsilon)&=&\prod_{k<r}(\varepsilon_r-\varepsilon_k)\,\,=\,\,\sum_{\sigma\in S_N}{\rm sgn}(\sigma)\varepsilon^0_{\sigma(1)}\varepsilon^1_{\sigma(2)}\cdots\varepsilon^{N-1}_{\sigma(N)}\label{VdM}
\eea
\bea
\bar{\Delta}(\varepsilon)&=&\prod_{k<r}(\varepsilon_r^{-1}-\varepsilon_k^{-1})\,\,=\,\,\sum_{\tau\in S_N}{\rm sgn}(\tau)\varepsilon^0_{\tau(1)}\varepsilon^{-1}_{\tau(2)}\cdots\varepsilon^{-N+1}_{\tau(N)}\label{barVdM}
\eea
Using the above results, the partition function becomes
\bea
Z(x^i,y^i,z^I)&=&{1\over (N!)^2 (2\pi i)^{2N}}\,\,\oint \prod_{j=1}^N {d\varepsilon_{1j}\over\varepsilon_{1j}}\,\,\Delta(\varepsilon_1)\bar{\Delta}(\varepsilon_1)\,\,\oint \prod_{j=1}^N {d\varepsilon_{2j}\over\varepsilon_{2j}}\,\,\Delta(\varepsilon_2)\bar{\Delta}(\varepsilon_2)\cr\cr
&&\times \prod_{I_1=1}^{b_1}{1\over (1-z_1^{I_1})^N}\prod_{1\le k<r\le N}{1\over (1-{\varepsilon_{1r}\over\varepsilon_{1k}}z_1^{I_1})(1-{\varepsilon_{1k}\over\varepsilon_{1r}}z_1^{I_1})}\cr\cr
&&\times \prod_{I_2=1}^{b_2}{1\over (1-z_2^{I_2})^N}\prod_{1\le s<t\le N}{1\over (1-{\varepsilon_{2t}\over\varepsilon_{2s}}z_2^{I_2})(1-{\varepsilon_{2s}\over\varepsilon_{2t}}z_2^{I_2})}\cr\cr
&&\times\prod_{A_1=1}^f \prod_{1\le i_1,j_1\le N}
{1\over (1-w_{12}^{A_1}{\varepsilon_{1i_1}\over\varepsilon_{2j_1}})}
\prod_{A_2=1}^f \prod_{1\le i_2,j_2\le N}
{1\over (1-w_{21}^{A_2}{\varepsilon_{2i_2}\over\varepsilon_{1j_2}})}
\eea
We now simplify the integrand of the above formula. According to (\ref{VdM}) and (\ref{barVdM}) each of the products $\Delta\bar{\Delta}$ involves a sum over permutations $\sigma$ and $\tau$. For each distinct term in the sum over $\sigma$, consider the distinct change of variables given by a permutation of the eigenvalues $\varepsilon_{1,i}\to \varepsilon_{1,\sigma(i)}$, $\varepsilon_{2,i}\to \varepsilon_{2,\sigma(i)}$. All factors in the integrand are invariant under this change except for the Van der Monde determinants which are both either invariant or both pick up a minus sign. Consequently each product $\Delta\bar{\Delta}$ is invariant. The net result of this change of variables is that the sum over $\sigma$ can be performed and we find
\bea
\Delta(\varepsilon)\bar{\Delta}(\varepsilon)\to  N!\varepsilon^0_1\varepsilon^1_2\cdots\varepsilon^{N-1}_{N}
\prod_{k<r}(\varepsilon_r^{-1}-\varepsilon_k^{-1})
\eea
It is now useful to change variables to the new variables\footnote{So there will be two sets of new variables $\varepsilon_{1j}=t_{11}t_{12}\cdots t_{1j}$ and $\varepsilon_{2j}=t_{21}t_{22}\cdots t_{2j}$.} $t_j$ defined by $\varepsilon_j=t_1t_2\cdots t_j$. A straightforward computation proves that
\bea
N!\varepsilon^0_1\varepsilon^1_2\cdots\varepsilon^{N-1}_{N}
\prod_{k<r}(\varepsilon_r^{-1}-\varepsilon_k^{-1})&=&N!\prod_{2\le k\le r\le N}(1-t_{k,r})
\eea
and
\bea
\prod_{1\le k<r\le N}(1-{\varepsilon_r\over\varepsilon_k}z)(1-{\varepsilon_k\over\varepsilon_r}z)&=&\prod_{2\le k\le r\le N}(1-zt_{k,r})(1-zt_{k,r}^{-1})
\eea
where\footnote{Again we will have $t_{1k,1r}=t_{1k}t_{1\,k+1}\cdots t_{1\,r-1}t_{1r}$ and $t_{2k,2r}=t_{2k}t_{2\,k+1}\cdots t_{2\,r-1}t_{2r}$.} $t_{k,r}=t_kt_{k+1}\cdots t_{r-1}t_r$. The Jacobian of this transformation is
\bea
J=\det {\partial\varepsilon_i\over \partial t_j}=\prod_{j=1}^N {\varepsilon_j\over t_j}
\eea
Thus, the partition function finally becomes
\bea
Z(x^i,y^i,z^I)&=&\prod_{I_1=1}^{b_1}{1\over (1-z_1^{I_1})^N} \prod_{I_2=1}^{b_2}{1\over (1-z_2^{I_2})^N} {1\over (2\pi i)^{2N}}\oint \prod_{j=1}^N {dt_{1j}\over t_{1j}}\,\oint \prod_{j=1}^N {dt_{2j}\over t_{2j}}\cr \cr
&& \times \prod_{2\le k\le r\le N}\frac{(1-t_{1k,1r})(1-t_{2k,2r})}{f_{k,r}}\,  \prod_{A_1=1}^f \prod_{1\le i_1,j_1\le N}
{1\over (1-w_{12}^{A_1}{t_{11,1i_1}\over t_{21,2j_1}})} \cr\cr
&& \times \prod_{A_2=1}^f \prod_{1\le i_2,j_2\le N}
{1\over (1-w_{21}^{A_2}{t_{21,2i_2}\over t_{11,1j_2}})}
\label{finalPF}
\eea
where
\bea
f_{k,r}&=&\prod_{I_1=1}^{b_1}\prod_{I_2=1}^{b_2}\,(1-z_1^{I_1} t_{1k,1r})(1-z_1^{I_1} t_{1k,1r}^{-1})(1-z_2^{I_2} t_{2k,2r})(1-z_2^{I_2} t_{2k,2r}^{-1})\cr\cr
&&
\eea
After fixing a specific value of $N$ in (\ref{finalPF}), the resulting expression is used to evaluate the exact partition function. Each integral over the variables $t_{1i},t_{2i}$ is performed along the unit circle in the complex plane and can be evaluated using the residue theorem. 

\section{Some Partition Functions}

\subsection{$f=1$ and $b_1=0=b_2$}

The new ingredient in the partition functions we compute here is the presence of bifundamental fields. To explore this we can set  $b_1=0=b_2$. The simplest case to study then is $f=1$, i.e. a single species of each type of fundamental field, say $\Gamma_{12}$ and $\Gamma_{21}$. The reason why it is useful to explore this specific case is that it is simple enough that we can compute the expected answer for the Molien-Weyl partition function in a second way. Since we must have  $\Gamma_{12}$ and $\Gamma_{21}$ alternating in the trace the complete set of invariants is given by
\bea
\Tr \left((\Gamma_{12}\Gamma_{21})^n\right)
\eea
This is the space of invariants of a single matrix $M=\Gamma_{12}\Gamma_{21}$. Thus, for the theory with $U(N)\times U(N)$ symmetry we expect the Molien-Weyl partition function must take the form
\bea
Z(w_{12},w_{21})&=&\prod_{n=1}^N {1\over 1-(w_{12}w_{21})^n}
\eea
We can easily test this expectation for some values of $N$:

{\vskip 0.5cm}

\noindent
{\bf N=2:} The Molien-Weyl partition function is
\bea
Z(w_{12},w_{21})&=&{1\over (2\pi i)^4}\oint {dt_{11}\over t_{11}}\oint {dt_{12}\over t_{12}}\oint {dt_{21}\over t_{21}}\oint {dt_{22}\over t_{22}}\,\,\frac{(1-t_{12})(1-t_{22})}{(1 - {t_{11}w_{12}\over t_{21}}) 
(1-{t_{11}t_{12}w_{12}\over t_{21}})}\cr\cr
&&\times\frac{1}{(1-{t_{11}w_{12}\over t_{21}t_{22}})(1-{t_{11}t_{12}w_{12}\over t_{21}t_{22}})(1-{t_{21}w_{21}\over t_{11}})(1-{t_{21}w_{21}\over t_{11}t_{12}})}\cr\cr
&&\times\frac{1}{(1-{t_{21}t_{22}w_{21}\over t_{11}})(1-{t_{21}t_{22}w_{21}\over t_{11} t_{12}})}
\eea
A straightforward (repeated) application of the residue theorem then leads to
\bea
Z(w_{12},w_{21})&=&{1\over (1-w_{12}w_{21})(1-(w_{12}w_{21})^2)}
\eea

{\vskip 0.5cm}

\noindent
{\bf N=3:} In this case there are 6 nested contour integrations. The result of the contour integration is
\bea
Z(w_{12},w_{21})&=&{1\over (1-w_{12}w_{21})(1-(w_{12}w_{21})^2)(1-(w_{12}w_{21})^3)}
\eea

{\vskip 0.5cm}

\noindent
{\bf N=4:} In this case there are 8 nested contour integrations. The result of the contour integration is
\bea
Z(w_{12},w_{21})&=&{1\over (1-w_{12}w_{21})(1-(w_{12}w_{21})^2)(1-(w_{12}w_{21})^3)(1-(w_{12}w_{21})^4)}
\eea

\subsection{$N=2$, $f=1$ and $b_1=1=b_2$}
One can also study the partition function for a pair of bosonic matrices (represented by the variables $z_{1,2}$) in the adjoint representation of each $U(N)$ of $U(N) \times U(N)$, as well as a pair of bifundamentals. For simplicity, consider the $N=2$ theory. The partition function is easily obtained by evaluating (\ref{finalPF}) as
\bea
Z &=& \frac{1}{(1- w_{12} w_{21})(1- w_{12}^2 w_{21}^2)} \times \frac{1}{(1-z_1)(1-z_1^2)(1-z_2)(1-z_2^2)} \cr \cr
&& \times \frac{1+ w_{12}^2 w_{21}^2 z_1 z_2}{(1-w_{12} w_{21} z_1) (1-w_{12} w_{21} z_2) (1-w_{12} w_{21} z_1 z_2)}
\eea
Notice that the Hironaka decomposition is valid with the finer grading that we are using.

\subsection{$N=2$ and $b_1=0=b_2$}

We can also consider the partition function for a variable number of bifundamental fields in the absence of matrices.  For the sake of simplicity (and conciseness) we will consider only the parition function that is ungraded in the various species of bifundamentals so that $w_{12}^{A_1} = w_{12}$ and $w_{21}^{A_2} = w_{21}$.  The first few examples are given by 
\begin{eqnarray}
Z_{f=1} &=& \frac{1}{(1 - w_{12}w_{21})(1 - (w_{12}w_{21} )^2 )}    \nonumber \\ 
Z_{f=2} &=& \frac{1}{(1 - w_{12}w_{21})^8(1 - (w_{12}w_{21} )^2 )^9}\left(1 + 5 (w_{12}w_{21})^2 + 5 (w_{12}w_{21})^4 + (w_{12}w_{21})^6 \right)  \nonumber \\
 Z_{f=3} &=& \frac{1}{(1 - w_{12}w_{21})^8(1 - (w_{12}w_{21} )^2 )^9}\left(1 +  (w_{12}w_{21}) + 37 (w_{12}w_{21})^2 + 56 (w_{12}w_{21})^3\right.    \nonumber \\
&&\quad \left. + 353 (w_{12}w_{21})^4 + 389 (w_{12}w_{21})^5 + 1037 (w_{12}w_{21})^6 + 704 (w_{12}w_{21})^7 \right.   \nonumber \\
&&\quad + 1037(w_{12}w_{21})^8 + 389 (w_{12}w_{21})^9 + 353(w_{12}w_{21})^{10} + 56(w_{12}w_{21})^{11}  \nonumber \\
&&\quad\left. + 37(w_{12}w_{21})^{12} + (w_{12}w_{21})^{13} + (w_{12}w_{21})^{14} \right)
\end{eqnarray}

As discussed in the main text, the total number of primary invariants is given by $8 (f-1) + 1$.  For this case we can do even better and we find that, for $f>1$, there are $4(f-1)$ primaries consisting of one pair of $\Gamma^A$, $\bar{\Gamma}^B$ bifundamentals and  $4(f-1)+1$ primaries consisting of two pairs of $\Gamma^A$, $\bar{\Gamma}^B$ bifundamentals.  The number of secondary invariants grows exponentially with $f$ as can be seen from Table \ref{N2Table}).

\begin{table}[h]
\begin{center}
\begin{tabular}{|c|c|}
\hline 
f & number of secondary invariants \\
\hline
2 & 12   \\
\hline 
3 & 4 452 \\
\hline
4 & 4 393 312 \\
\hline 
5 & 6 525 650 100 \\
\hline
6 & 12 157 761 241 248 \\
\hline 
7 & 26 159 945 623 505 568 \\
\hline 
8 & 62 184 870 388 103 240 448 \\
\hline 
9 & 159 011 509 481 010 770 452 308 \\
\hline
\end{tabular}
\end{center}
\caption{The number of secondary invariants as a function of the number of species of bifundamentals for $N=2$ and $b_1 = 0 = b_2$}
\label{N2Table}
\end{table}

\subsection{$N=3$ and $b_1=0=b_2$}

Turning now to $N=3$ the expressions for the partition functions become more involved.  As with the $N=2$ case we are able to extract more detail of the structure of primary invariants.  The total number of primary invariants is given by $18 (f-1) + 1$ of which (for $f>0$) we find that $3(f-1)$ consist of one pair of $\Gamma^A$, $\bar{\Gamma}^B$ bifundamentals, $9(f-1)$ consist of two pairs of $\Gamma^A$, $\bar{\Gamma}^B$ bifundamentals and $6(f-1)+1$ consist of three pairs of $\Gamma^A$, $\bar{\Gamma}^B$ bifundamentals.  The number of secondary invariants grow rapidly with $f$ as can be seen from Table \ref{N3Table}.
\begin{table}[h]
\begin{center}
\begin{tabular}{|c|c|}
\hline 
f & number of secondary invariants \\
\hline
2 & 11 784   \\
\hline 
3 & 3 644 220 637 440 \\
\hline
4 & 11 121 748 209 170 781 930 624 \\
\hline 
5 & 91 233 439 246 830 531 648 669 228 453 600 \\
\hline
6 & 1 300 304 572 796 023 662 859 714 921 626 814 097 931 264 \\
\hline
\end{tabular}
\end{center}
\caption{The number of secondary invariants as a function of the number of species of bifundamentals for $N=3$ and $b_1 = 0 = b_2$}
\label{N3Table}
\end{table}

\section{Secondary Invariants at $N=2$}

Although we are not able to compute the number of secondary invariants defined in a multi-matrix system of $d$ matrices analytically at generic values of $N$, for $N=2$ we have been able to obtain a formula. Some example partition functions include (the subscript on $Z_d(x)$ is the value of $d$)
\bea
Z_1(x)&=&\frac{1}{(1-x) \left(1-x^2\right)}\qquad\qquad\qquad\qquad
Z_2(x)\,\,=\,\,\frac{1}{(1-x)^2 \left(1-x^2\right)^3}\cr\cr
Z_3(x)&=&\frac{1+x^3}{(1-x)^3 \left(1-x^2\right)^6}\qquad\qquad\qquad\quad
Z_4(x)\,\,=\,\,\frac{1+x^2+4x^3+x^4+x^6}{(1-x)^4 \left(1-x^2\right)^9}\cr\cr
Z_5(x)&=&\frac{1+3 x^2+10 x^3+6 x^4+6 x^5+10 x^6+3 x^7+x^9}{(1-x)^5 \left(1-x^2\right)^{12}}\cr\cr
Z_6(x)&=&\frac{1+6 x^2+20 x^3+21 x^4+36 x^5+56 x^6+36 x^7+21 x^8+20 x^9+6 x^{10}+x^{12}}{(1-x)^6 \left(1-x^2\right)^{15}}\cr\cr
Z_7(x)&=&\frac{1}{(1-x)^7 \left(1-x^2\right)^{18}}\Big(1+10x^2+35x^3+55x^4+126x^5+220x^6+ 225x^7+225x^8\cr\cr
&&\quad +220x^9+126x^{10}+55x^{11}+35x^{12}+10 x^{13}+x^{15}\Big)
\eea
In this case there are $d$ primary invariants of length 1 and $3d-3$ primary invariants of length 2 so that the general form of the partition function is
\bea
Z_d(x)&=&\frac{P_d(x)}{(1-x)^d(1-x^2)^{3d-3}}
\eea
Notice that $d + 3d -3=1+(d-1) 2^2$ so that we have the correct number of primary invariants. The numerator $P_d(x)$ is always a palindromic polynomial. We have obtained the following formula for the number of secondary invariants
\bea
P_d(1)=\frac{2^{d-2} (2d-4)!}{(d-2)! (d-1)!}
\eea
In the large $d$ limit this behaves, at leading order in $d$, as
\bea
P_d(1)=e^{3d\log 2}
\eea
so that the leading entropy defined from this number of secondary invariants
\bea
\log P_d(1)=3d\log 2
\eea
If we study a single matrix field theory on a lattice, we would have a different species of matrix at each lattice site. Consequently, we could use the above formula to count the number of secondary invariants provided we set the number of flavours $d$ equal to the number of lattice sites. The volume of the field theory is, up to a multiplicative constant, given by the total number of sites. Thus, the above formula would lead to an extensive entropy, which is satisfying.
  
\end{appendix}

\end{document}